\documentclass{elsart}

\usepackage{natbib}

\usepackage{graphicx}
\DeclareGraphicsExtensions{.ps}
\usepackage{subfigure}

\usepackage{amssymb}

\begin{document}

\begin{frontmatter}

\title{INTEGRAL/IBIS search for e$^{-}$e$^{+}$ annihilation radiation from the
Galactic Center Region}

\author[iasf]{G. De Cesare}
\author[iasf]{A. Bazzano}
\author[iasf]{F. Capitanio}
\author[iasf]{M. Del Santo}
\author[cesr]{V. Lonjou}
\author[iasf]{L. Natalucci}
\author[iasf]{P. Ubertini}
\author[cesr]{P. Von Ballmoos}

\address[iasf]{Istituto di Astrofisica Spaziale e Fisica cosmica
   (IASF) - CNR, via Fosso del Cavaliere 100, 00133 Roma, Italy}
\address[cesr]{Centre d'Etude Spatiale des Raynnements, CNRS/UPS,
   B.P. 4346, 31028 Toulouse Cedex 4, France}

\begin{abstract}
Electron-positron annihilation radiation from the Galactic Center region
has been detected since the seventies, but its astrophysical
origin is still a topic of a scientific debate. We have analyzed data of
the gamma-ray imager IBIS/ISGRI onboard of ESA's INTEGRAL platform in the
e$^{-}$e$^{+}$ line. During the first year of the missions Galactic Center
Deep Exposure no evidence for point sources at 511 keV has been found in
the ISGRI data; the $2 \sigma$ upper limit for resolved single
point sources is estimated to be $1.6\times 10^{-4}\ ph\ cm^{-2}\ s^{-1}$.

\end{abstract}


\end{frontmatter}

\section{Introduction}
In the seventies, balloon instruments provided first evidence for
e$^{-}$e$^{+}$ annihilation from the Galactic Center region.  As the line
was discovered at an energy of 476 $\pm$ 26 keV \citep{Johnson1972}, the
physical process behind the emission was initially ambiguous and had to
await for the advent of high resolution spectrometers.  In 1977, germanium
semiconductors, flown for the first time on balloons, allowed to establish
the identification of the narrow annihilation line at 511 keV, its width
turned out to be of a few keV only (Albernhe {\it et al.} 1981, Leventhal
{\it et al.} 1978).  The eighties were marked by ups and downs in the
measured 511 keV flux through a series of observations performed by the
balloon-borne germanium detectors (principally the telescopes of
Bell-Sandia and GSFC).  The fluctuating results were interpreted as the
signature of a compact source of annihilation radiation at the Galactic
Center (see e.g.  Leventhal 1991).  Additional evidence for this scenario
came initially from HEAO-3 (Riegler {\it et al.} 1981) reporting
variability in the period between fall 1979 and spring 1980.  Yet, during
the early nineties, this interpretation was more and more questioned, since
neither eight years of SMM data (Share {\it et al.} 1990) nor the revisited
data of the HEAO-3 Ge detectors (Mahoney {\it et al.} 1993) showed evidence
for variability in the 511 keV flux.  Throughout the nineties, CGRO's
Oriented Scintillation Spectrometer Experiment (OSSE) measured steady
fluxes from a galactic bulge and disk component (Purcell {\it et al.} 1997)
and rough skymaps became available based on data from OSSE, SMM and TGRS.
A possible third component at positive Galactic latitude  
was attributed to an annihilation fountain in the Galactic center.

Amongst the different models that have been proposed to explain the origin
of positrons in the GC let us only mention a few : radioactive nuclei
produced by nucleosynthesis \citep{Ram1979}, neutron stars or black holes
\citep{Ling1983}, pulsars \citep{Stur1971}, cosmic ray interaction with the
interstellar medium \citep{Koz1987}, gamma-ray bursts \citep{Ling1984} and
light dark matter physics \citep{Boehm2004}.

Since the launch of INTEGRAL in October 2002, ample observing time of the
mission's core program has been devoted to the Galactic Center Deep
Exposure (GCDE).  A map obtained by SPI during the first year GCDE shows an
extended 511 keV emission $(8^{+3}_{-2}\deg FWHM)$ centered symmetrically
on the Galactic Center (Jean {\it et al.} 2004, Weidenspointner {\it et
al.} 2004).  The corresponding bulge flux is $0.96^{+0.21}_{-0.14}\times
10^{-3}\ ph\ cm^{-2}\ s^{-1}$ , with the uncertainty being dominated by the
width of the Gaussian intensity distribution.  Spectroscopy of 511 keV line
emission from the bulge resulted in a best fit energy of
$511.02^{+0.08}_{-0.09}$ keV and an intrinsic line width of
$2.67^{+0.30}_{-0.33}$ keV FWHM (Lonjou {\it et al.} 2004).

SPI has not detected emission from positive latitudes, and the Galactic
plane emission - if existent - must be on a lower level than reported by
OSSE.  The GC emission at 511 keV can not be explained by a single source,
yet the contribution of a number of point sources can not be excluded.  The
simple bulge morphology of the 511 keV emission observed by SPI is
suggestive for a e$^{+}$ origin in the Galaxie's old stellar population.

In the future, additional exposure and improved knowledge of background
systematic will refine SPI's image of Galactic e$^{-}$e$^{+}$
annihilation, and better constrain the many models proposed for its origin.

Also, an important contribution for constraining the models is expected to
come from INTEGRAL's imager IBIS. With its superior angular resolution and
good sensitivity for point sources, IBIS will help to decide whether the
extended bulge emission is of genuinely diffuse origin or the result of
a number of blended compact sources.
In this article we present preliminary results of an ISGRI data analysis in
the 511 keV band during the first year of the missions Galactic Center Deep
Exposure.  This analysis uses the IBIS Standard Analysis pipeline and is
therefore optimized for - and limited to - point sources detection. 
For the diffuse emission detection, a so called  ``light bucket'' 
data analysis technique has been performed for SPI 511 keV data analysis and 
at lower energies for IBIS/ISGRI \citep{Lebrun2003n}. The light
bucket analysis at 511 keV for IBIS is in progress and will be 
presented elsewhere.

\begin{figure} [t]
\begin{center}
\end{center}
\includegraphics[width=12cm, height=8cm, bb = 2cm 12cm 20cm 26cm,clip=true]{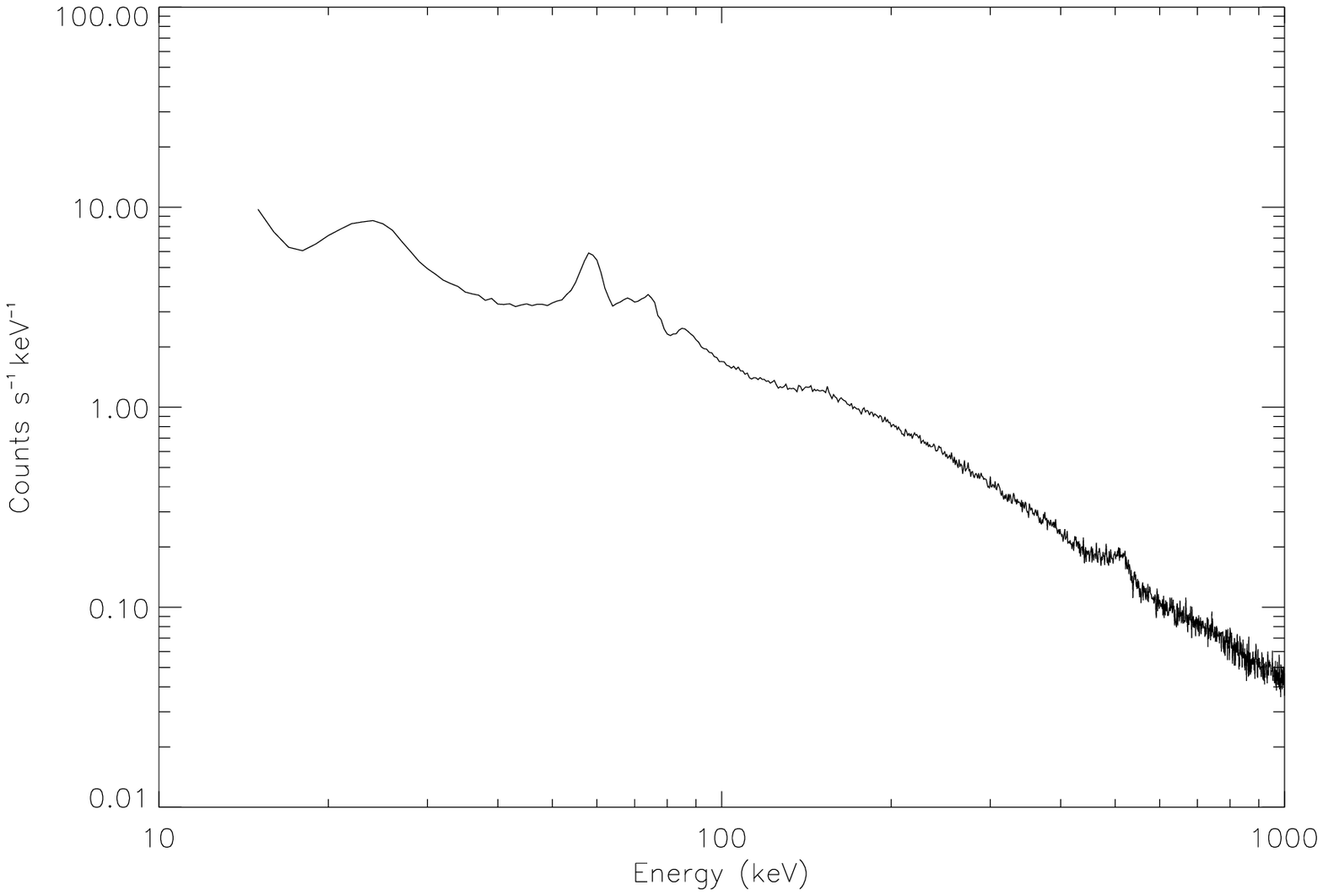}
\caption{ISGRI Background Spectrum obtained from a science window sample
   (acquired during the revolution 53); the exposure time is 1800 seconds.
   The 511 keV line component is clearly
   visible. The lines below 100 keV are due to lead and tungsten
   fluorescence in the telescope.}
\label{fig:bkg}
\end{figure}

\section{Data analysis}
The IBIS instrument \citep{Ubertini2003} on board of the INTEGRAL
satellite \citep{Winkler2003} is an X and Gamma-ray
coded mask telescope  with a large field of view
($29^{\circ} \times 29^{\circ}$) and a good angular resolution
($12^{'}$). IBIS has also spectral capabilities in the wide
energy band from 15 keV to 10 MeV, with
a reasonable energy resolution. The position sensitive
detector is based on two layer, ISGRI \citep{Lebrun2003},
a CdTe pixelated low energy $128 \times 128$ matrix, covering the range
from 15 keV to 1 MeV, and PICsIT \citep{Labanti2003}, a CsI
  $64 \times 64$ matrix, covering the range from 175 keV to 10 MeV.
The IBIS detectors Quantum Efficiencies (QE)
\citep{Decesare2001} and the background count rate depend
on the photon energy. The ISGRI
Quantum Efficiency (QE) is equal to 2.0 \% at 511 keV, due to the 
low thickness (2 mm) of the CdTe detectors,
but the background rate (fig. \ref{fig:bkg}) is quite low.
Below 100 keV the ISGRI background spectrum show the emission lines due
to the lead and tungsten fluorescence photons generated
in the IBIS telescope. The 511 keV line,  mainly caused by
the hight energy particles interaction with the INTEGRAL materials,
is also present.

The IBIS standard analysis
\citep{Goldwurm2003} is optimized for detection and spectral
extraction of point sources, also in crowded region.
The IBIS data and the  Off-line Scientific Analysis
(OSA) software are delivered by the Integral Science Data Center
(ISDC) \citep{Courvoisier2003}.

\section{Results}

\begin{figure} [t]
\subfigure{\includegraphics[width=6.8cm,height=6.8cm]{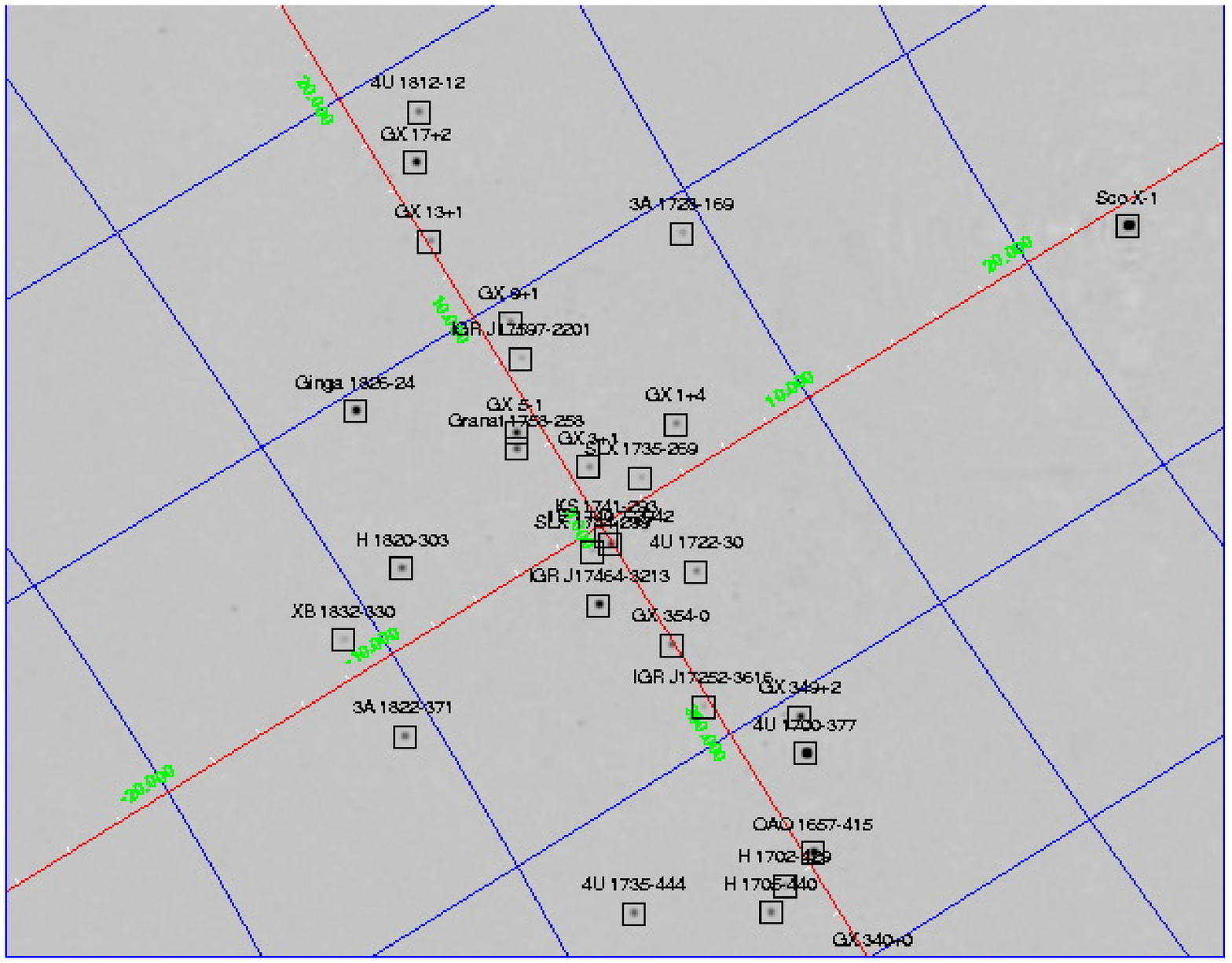}}
\subfigure{\includegraphics[width=6.8cm,height=6.8cm]{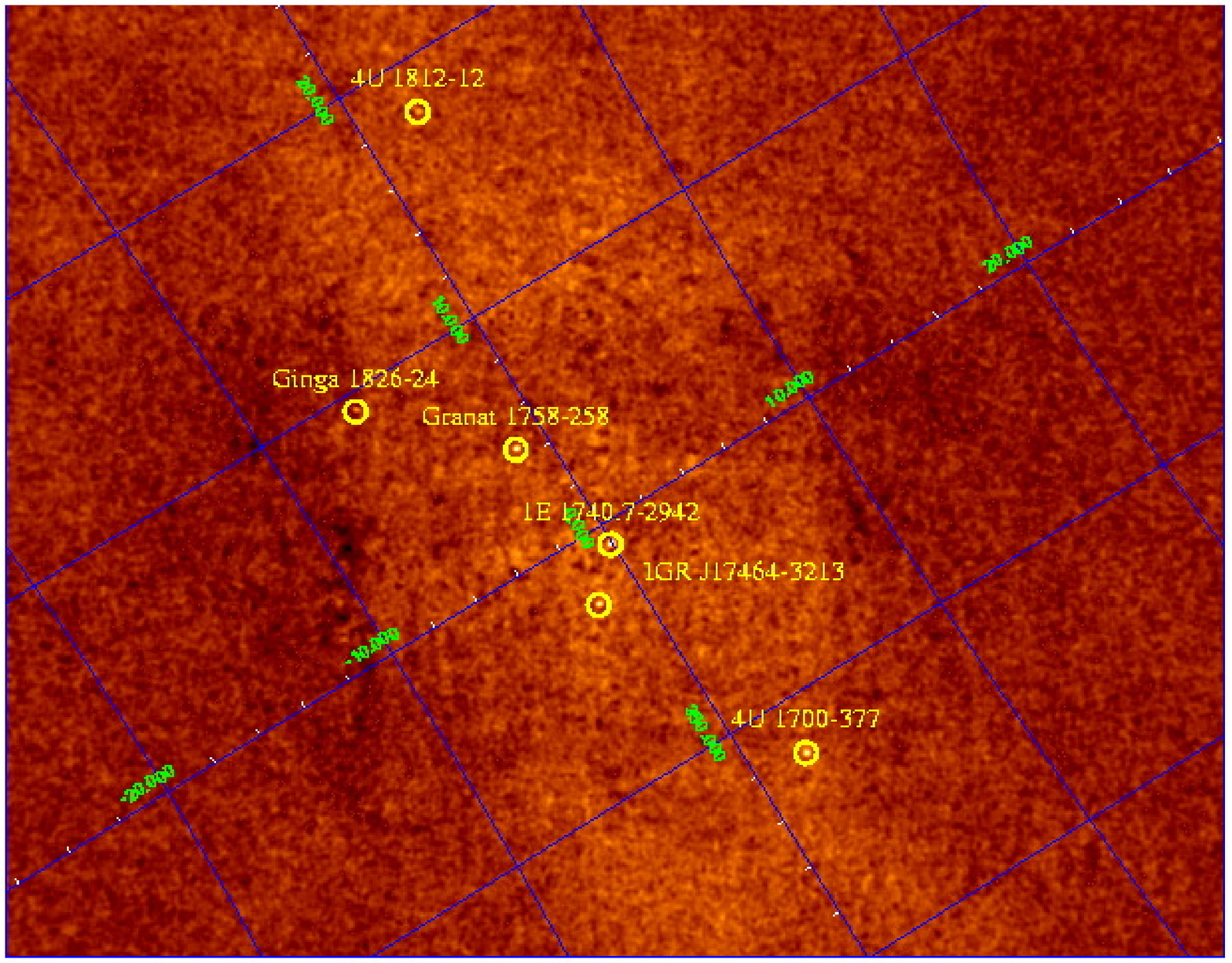}}
\caption{Two IBIS/ISGRI mosaic images of the Galactic Center Region
   in different energy bands.
   The low energy map on the left shows the mosaic in the energy band
   18 keV - 40 keV.
   The image on the right shows the ISGRI mosaic
   in the energy band 120 keV - 250 keV. At high energies, six
   sources are clearly detected: 2 NS LMXBs (Ginga 1826-24, 4U 1812-12),
   3 BH LMXBs (Granat 1758-258, 1E 1740.7-2942, IGR J17464-3213),
   1 NS or BH HMXB (4U 1700-377)}
\label{fig:mosaic}
\end{figure}

Eight ISGRI mosaic images have been obtained, from revolutions 46
to 123 in the following
energy intervals: [18 keV, 40 keV], [40 keV, 60 keV], [60 keV, 120 keV],
[120 keV, 250 keV], [250 keV, 435 keV], [435 keV, 485 keV],
[485 keV, 535 keV], [535 keV, 585 keV].
While at low energies the GC is a very crowded region, only
six hard X-ray sources are detected in the 120 keV - 250 keV energy band
(fig. \ref{fig:mosaic}). Some of these sources could in principle give a
contribute at 511 keV.

\begin{figure} [t]
\includegraphics[width=6.8cm,height=6.8cm]{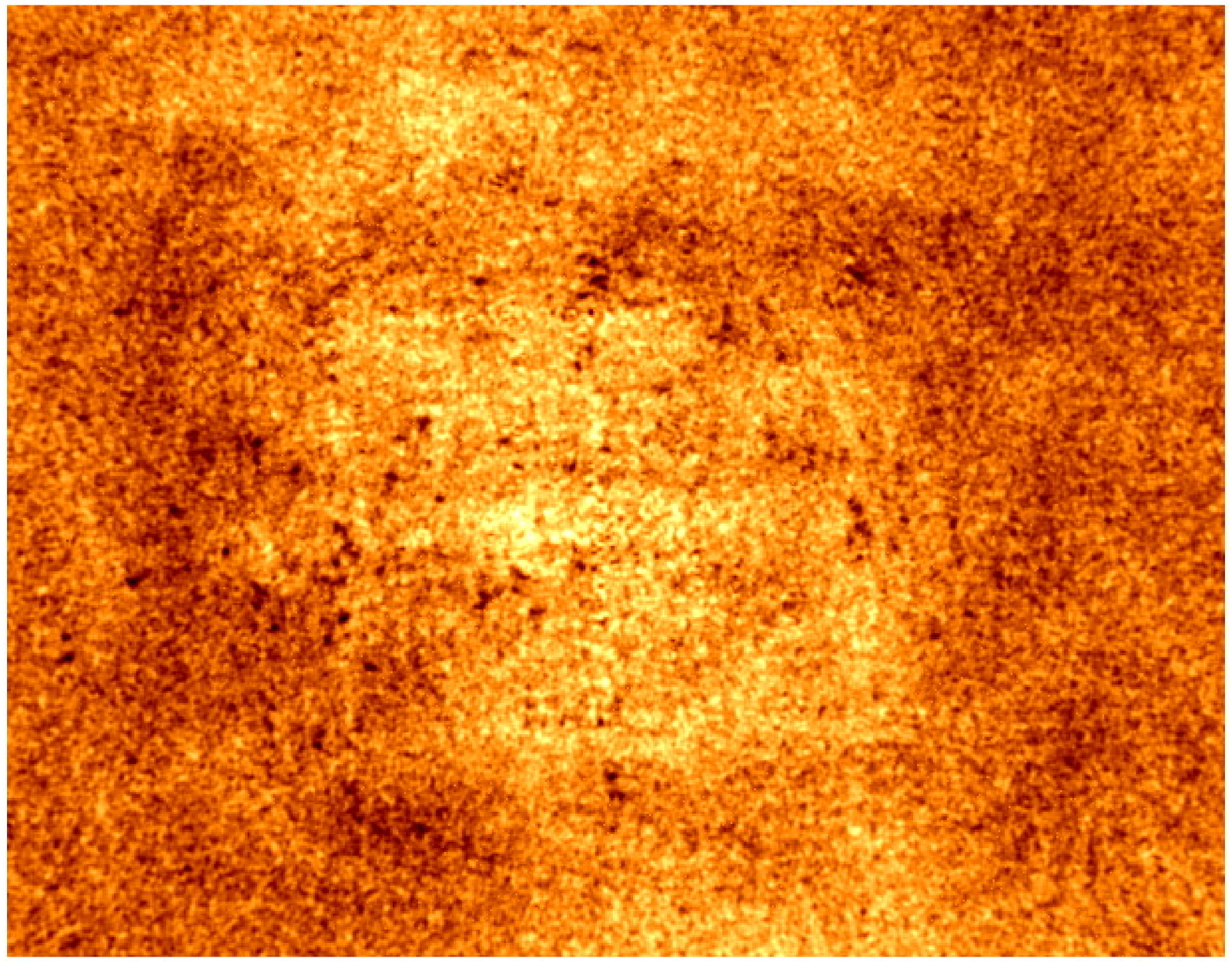}
\subfigure{\includegraphics[width=6.8cm,height=6.8cm]{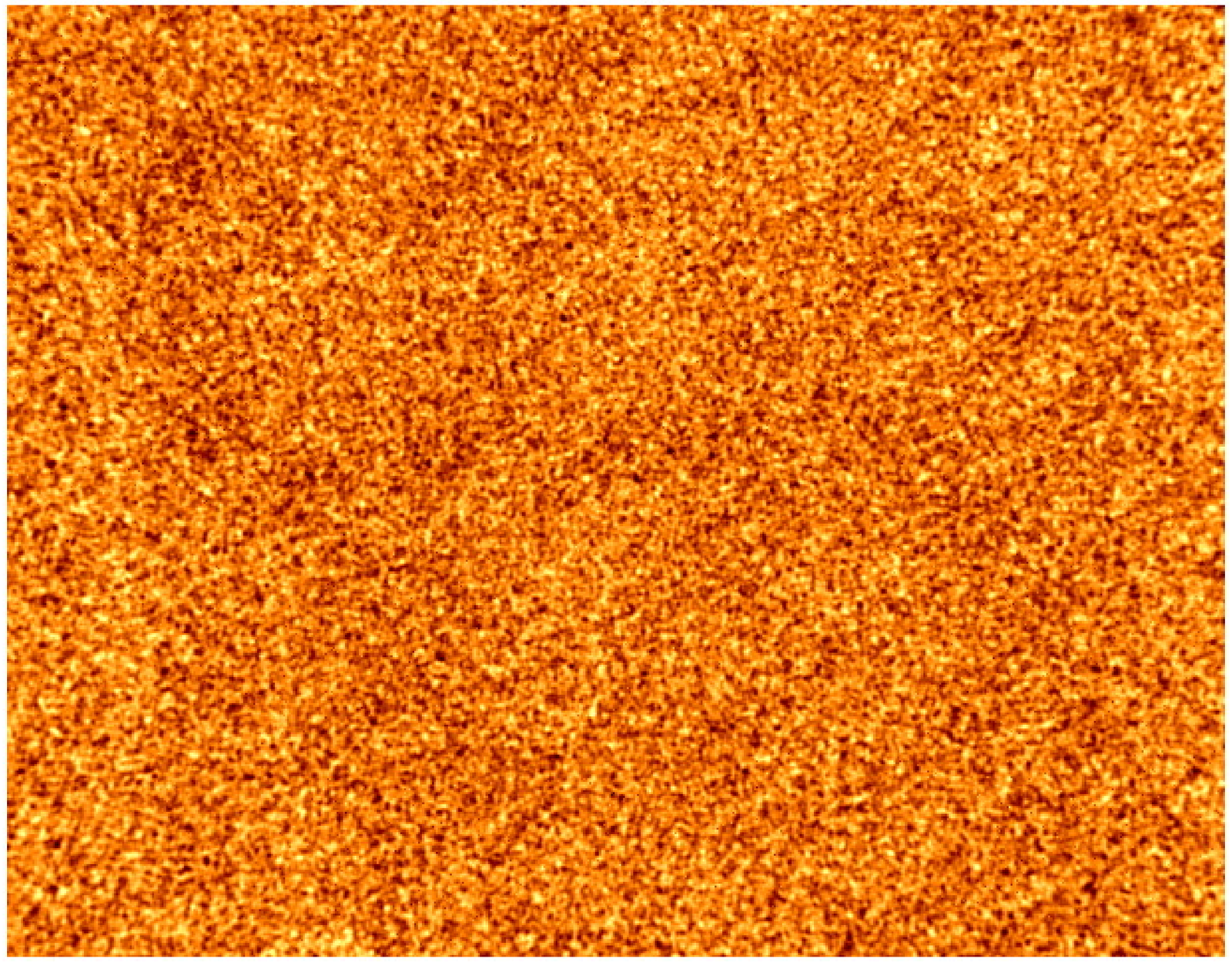}}
\caption{The standard processing of the GCDE ISGRI data has
not shown any evidence for 511 keV point sources.
The  image on the left shows the mosaic in the energy
band 485 keV - 535 keV.
The large structure that appears is due to
the overlap of the dithering points. After the correction
for these effects (image on the right panel),
the systematic structures in the image are strongly reduced}
\label{fig:511}
\end{figure}

The figure \ref{fig:511} show the IBIS/ISGRI mosaic at 511 keV, 
in the energy band 485-535 keV, with an
exposure time equal to 1.5 Msec in a large part of the Galactic Center
Region. The left panel in the figure show the mosaic image
as obtained by the standard pipeline without any correction;
the large structure in the image is mainly due to the overlapping
of the dithering pointings. After the subtraction of the
mean value of the maps in the adjacent bands (435-485 keV and 535-585 keV), 
this systematic noise is reduced from 36 \% to 8 \%, but the 
statistic noise is increased by a square root of 2 factor.

Despite the deep exposure, no 511 keV point sources have
been found. Starting from the ISGRI sensitivity that we have
estimated from the Crab calibration data
($2.9\times 10^{-4}\ ph\ cm^{-2}\ s^{-1}$, at 3 $\sigma$ for an exposure
time of 1 Msec), we obtain the flux upper limit 
$1.6\times 10^{-4}\ ph\ cm^{-2}\ s^{-1}$, at $2 \sigma$ of confidence 
level for an exposure time equal to 1.5 Msec.

\section{Conclusions}
At the present state of the IBIS/ISGRI GCDE data analysis we have
not found any evidence for 511 keV point sources in the Galactic Center
Region. Taking in account of the ISGRI sensitivity, the 
data set an upper limit (at $2 \sigma$ confidence level) of 
$1.6\times 10^{-4}\ ph\ cm^{-2}\ s^{-1}$ on the 511 keV flux for any
point source in the Galactic Center region.
The sensitivity improvement provided by the PICsIT data analysis 
and by the analysis of a more extended data set will give more costraints
on the 511 keV flux, or we will found a sample of new sources.

For a diffuse radiation the IBIS standard pipeline is not
appropriate. Dedicated data analysis strategy and methods
for this purpose are in progress.

\end{document}